\documentclass[journal]{IEEEtran}
\ifCLASSINFOpdf \else \fi

\usepackage{graphicx}
\usepackage[cmex10]{amsmath}
\usepackage{amssymb}
\usepackage{bm}
\usepackage{array}
\usepackage{url}
\usepackage[noadjust]{cite}

\begin{document}

\title{Efficient calculation of inductive coupling for arrays of wire ring resonators}

\author{{S. K. Zhogolev$^1$, M. Y. Petrov$^1$, A. O. Makarenko$^1$, M. Lapine$^{1,2}$, and A. A. Shcherbakov$^1$ \\ $^1$School of Physics and Engineering, ITMO University, St-Petersburg, Russia \\ $^2$Physics Department, Lomonosov Moscow State University, Moscow 119991, Russia}
}


\maketitle

\begin{abstract}
Generalization of the inductance to the case of non-quasistatic electromagnetic field oscillations appears to be fruitful when considering wireless power transfer and RF metamaterials consisting of thin wire loop meta-atoms. When dealing with large systems of interacting loops carrying currents, efficiency and precision of calculation in the presence of retardation is crucial. In this work, we derive a series expansion of such generalized inductance and propose a way for its efficient numerical approximation. Illustrative examples are provided both for inductance convergence of a pair of two loops and extinction efficiency for scattering by metamaterial samples.
\end{abstract}

\begin{IEEEkeywords}
wire loop resonator, inductive coupling, metamaterial, inductance.
\end{IEEEkeywords}

\IEEEpeerreviewmaketitle

\section{Introduction}\label{sec:intro}

Inductive coupling is an important phenomenon whose applications range from traditional power transformers and magnetic lines to the actively developing area of wireless power transfer (WPT) \cite{Jayalath2021,Kim2026,Okasili2022}. Owing to its indisputable practical importance, the phenomenon lies behind the fundamental physical behavior of a range of artificial structures, namely metamaterials and metasurfaces, composed of various kinds of shaped wires and split-ring resonators \cite{Simovski2012,Duran2012}.

Calculation of mutual inductance is a very early and fundamental problem in electrodynamics, and in general it obeys a well-established procedure whether in theoretical physics \cite{Landau,Stratton1941} or electrical engineering \cite{poletkin2019efficient}. Most applications, however, either refer to near-field coupling, such as in low-frequency scenarios, as is customary with electric circuit design or magnetic resonance imaging, or, vice versa, far-field coupling such as that between loop antennas \cite{geogakopoulos1999coupling}.

Contemporary interest within a couple of the mentioned major research areas -- wireless power transfer (WPT) and metamaterials -- prompts a more general analysis covering also the intermediate scale range, whereby the distance between inductively coupled objects is comparable to the wavelength of the relevant radiation. Then, if for WPT applications the problem is relaxed somewhat through the possibility of full-wave numerical simulations for a given design, for metamaterials, involving thousands of structural elements, that would not constitute a practically feasible solution.


Indeed, recent progress in the calculation of the quasi-static electromagnetic response of metamaterials with a large number of elements, progressing from a few thousand \cite{LapJelFre10}, a few tens of thousands \cite{LapMcPPou16}, towards millions of meta-atoms \cite{makarenko2026electromagnetics}, calls for an extension of such calculations beyond the threshold of quasi-static analysis. To this end, it would be desirable to explore metamaterials where the size of a meta-atom is still much smaller than the wavelength, as required \cite{GorLapTre06}; however, their overall size becomes comparable to the operating wavelength. 

In this situation, retardation effects cannot be neglected, implying significant spatial dispersion \cite{Agranovich} at the level of theoretical description \cite{Sil7,BaeJelMar08,SilBaeJel09,Sim09}, and making it a challenge to perform direct precise calculations that take into account all the mutual interactions between meta-atoms within the entire structure \cite{LapJelMar10}. For the practical implementation of metamaterials and so-called 'metasurfaces', this aspect is quite important, as it is challenging to assemble large structures that fulfill all the requirements for a quasi-static consideration. The most sub-wavelength design for a meta-atom is a capacitively-loaded ring resonator \cite{MarJelFre11}, where a sufficiently large lumped capacitor may provide a resonance frequency corresponding to a wavelength a few orders of magnitude greater than the resonator size, so that even a large array of such resonators is still sufficiently smaller than that wavelength. Such a provision is not so easy to implement, and in practice it is more likely that the resonator size is only about 10-100 times smaller than the resonance wavelength. Under this assumption, mutual interaction between remote resonators within the structure, which is rather crucial for the overall electromagnetic response \cite{GorLapTre06}, must be calculated taking retardation into account.



Recently we have developed a quasistatic model for studying discrete metastructures composed of a huge number of meta-atoms \cite{makarenko2026electromagnetics}. In order to generalize our findings, in this work we explore ways for efficient calculation of non-quasistatic inductance and inductive coupling of large systems of ring resonators. Generalization of the mutual inductance of two current loops to the non-quasistatic case requires evaluating two contour integrals \cite{Zierhofer1996,Tatartschuk2012,Parise2020} (see the next section). This numerical procedure should be performed with sufficient accuracy and appears to be too resource consuming upon computation of an impedance or admittance matrix for hundreds of thousands and more resonators. Some analytical improvements were proposed in \cite{Parise2020,Parise2024} for cases of parallel circular loops. Here we extend these results and propose a new way for computation of approximate mutual inductance and admittance matrices for systems of coupled ring resonators. 

\section{Inductive coupling between two ring resonators}\label{sec:inductance}

\begin{figure}[h]
    \begin{center}
        \includegraphics[width=0.6\linewidth]{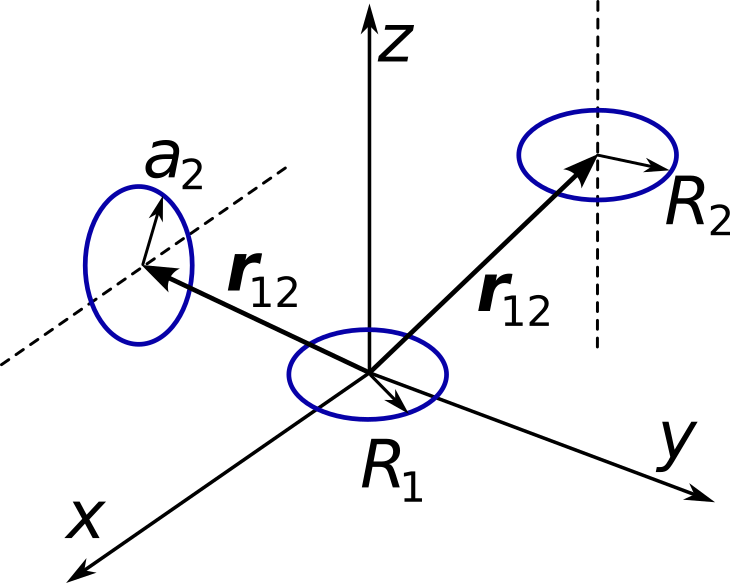}
        \caption{Schematic picture of two loop configurations analyzed in the paper: when the loops lie in parallel planes (central and right loops), and when the loops lie in orthogonal planes (central and left loops).}
        \label{fig:rings}
    \end{center}
\end{figure}

We focus on the problem of electromagnetic interaction of two circular loops of radii $a_1$ and $a_2$ and restrict the problem to cases when the loops lie either in parallel or orthogonal planes, as illustrated in Fig.~\ref{fig:rings}. This is the most widespread case for loop meta-atom layout. Coordinate vector connecting the loop centers is $\bm{R}=(\Delta x, \Delta y, \Delta z)^{\mathsf{T}}$. The main assumption for all further derivations is a negligible loop thickness. With this assumption, and taking the origin to coincide with the center of the first loop, all its points are described by the parameter $\phi_1$ such that $\bm{r}_1 = a_1(\cos\phi_1,\sin\phi_1,0)^{\mathsf{T}}$, while the points of the second loop are parametrized via $\phi_2$: $\bm{r}_2 = \bm{R} + a_2(\cos\phi_2,\sin\phi_2,0)^{\mathsf{T}}$. Also let us denote $\mathbf{r}_{12}=\mathbf{r}_{2}-\mathbf{r}_{1}$ and $r_{12}=|\mathbf{r}_{12}|$.

Here we consider time-harmonic fields and sources with the time dependence being expressed by the factor $\exp(-i\omega t)$. When a source current with amplitude $I_1$ flows in the first loop, an alternating current induced in the second loop is determined by an alternating magnetic flux crossing the loop. The retarded magnetic field of a current of the first loop can be expressed via the vector potential $\bm{A}$, which obeys the Helmholtz equation \cite{Stratton1941}
\begin{equation}
    \left(\nabla^2 + k^2\right)\bm{A} = -\mu_0 \bm{J}_1
    \label{eq:helm_A}
\end{equation}
under the Lorentz gauge, providing that the radiation occurs in vacuum. Here $k=\omega\sqrt{\varepsilon_0\mu_0}$ is the vacuum wavenumber, and $\varepsilon_0$, $\mu_0$ are the dielectric permittivity and magnetic permeability of vacuum. Under the assumption of a line source the solution of Eq.~(\ref{eq:helm_A}) has the form a closed loop integral
\begin{equation}
    \mathbf{A}(\mathbf{r}) = \frac{\mu_0 I_1}{4\pi} \ointop_{C_1} \frac{e^{ik|\mathbf{r} - \mathbf{r}_1|}}{|\mathbf{r} - \mathbf{r}_1|} d\mathbf{l}_1
    \label{eq:sol_A}
\end{equation}
so that the magnetic flux through the second loop reads
\begin{equation}
    \Phi_{12} = \mu_0 \intop_{S_2} \bm{H} \cdot d\bm{S}_2 = \ointop_{C_2} \mathbf{A} \cdot d\mathbf{l}_2
    \label{eq:flux_A}
\end{equation}

Following the conventional inductance definition in the quasi-static regime we write $\Phi_{12} = L_{12} I_1$,
where from
\begin{equation}
    \begin{split}
        L_{12} &= \frac{\mu_0}{4\pi} \intop_{S_2} d\bm{S}_2 \cdot \left[\, \ointop_{C_1} f(r_{12}) \, \mathbf{r}_{12}\times d\mathbf{l}_1 \right] = \\ &= \frac{\mu_0}{4\pi} \ointop_{C_2} \ointop_{C_1} \frac{e^{ikr_{12}}}{r_{12}} d\mathbf{l}_1 \cdot d\mathbf{l}_2
    \end{split}
    \label{eq:inductance_loops}
\end{equation}
which is consistent with previous research \cite{Zierhofer1996,Tatartschuk2012,Parise2020}. The function appearing in the first line explicitly reads
\begin{equation}
    f(r_{12}) = (ikr_{12}-1) \frac{e^{ikr_{12}}}{r_{12}^3}
    \label{eq:funcf}
\end{equation}
For further derivations we need also an expansion of this function around the point $R=|\mathbf{R}|$ up to the $1/R^2$ terms with explicitly extracted quasistatic term $f_{qs} = \left.f\right|_{k=0}$:
\begin{equation}
    \begin{split}
        f&(\mathbf{R}+\Delta\mathbf{r}) = (1-ikR)e^{ikR}f_{qs} + \\ & + \dfrac{e^{ikR}}{R^2} \sum_{n=1}^{\infty} \frac{(ik)^{n+1}}{n!R^n} (\mathbf{R}\cdot\Delta\mathbf{r})^n + o\left(\frac{1}{R^2}\right)
    \end{split}
    \label{eq:f_Taylor}
\end{equation}
Efficient calculation of the quasistatic term was described previously, e.g. see \cite{LapJelMar10,poletkin2019efficient}.

Recalling the special cases of loops lying either in parallel or orthogonal planes considered in this paper, we can simplify contour integration in Eq.~(\ref{eq:inductance_loops}) using the introduced angles $\phi_{1,2}$, which define points on the circular loops. For parallel circular loops the mutual inductance becomes
\begin{equation}
    \begin{split}
        L^{\|}_{12} &= \frac{\mu_0 a_1 a_2}{4\pi} \int_0^{2\pi} \!\!\! \int_0^{2\pi}  d\phi_1 d\phi_2 \cos(\phi_2-\phi_1) \times \\
        &\quad \times  \! \frac{e^{ik\sqrt{A^{\|} + B^{\|}\cos\phi_1 + C^{\|}\sin\phi_1}}}{\sqrt{A^{\|} + B^{\|}\cos\phi_1 + C^{\|}\sin\phi_1}} 
    \end{split}
    \label{eq:inductance_rings_P}
\end{equation}
where
\begin{equation}
    \begin{split}
        A^{\|}(\phi_2) =& \; \Delta z^2 + \Delta b^2 + 2a_2(\Delta x \cos\phi_2 + \Delta y \sin\phi_2) + \\ &+ a_1^2 + a_2^2 \\
        B^{\|}(\phi_2) =& \; -2\Delta x a_1 - 2a_1 a_2 \cos\phi_2 \\
        C^{\|}(\phi_2) =& \; -2\Delta y a_1 - 2a_1 a_2 \sin\phi_2 \\
    \end{split}
    \label{eq:inductance_ABC_P}
\end{equation}
and $\Delta b = \sqrt{(\Delta x)^2+(\Delta y)^2}$. For circular loops lying in orthogonal planes 
\begin{equation}
    \begin{split}
        L^{\perp}_{12} &= -\frac{\mu_0 a_1 a_2}{4\pi} \int_0^{2\pi} \!\!\! \int_0^{2\pi}  d\phi_1 d\phi_2 \cos\phi_1 \cos\phi_2 \times \\
        &\quad \times  \! \frac{e^{ik\sqrt{A^{\perp} + B^{\perp}\cos\phi_1 + C^{\perp}\sin\phi_1}}}{\sqrt{A^{\perp} + B^{\perp}\cos\phi_1 + C^{\perp}\sin\phi_1}} 
    \end{split}
    \label{eq:inductance_rings_O}
\end{equation}
where
\begin{equation}
    \begin{split}
        A^{\perp}(\phi_2) =& \; \Delta z^2 + \Delta b^2 + 2a_2(\Delta y \sin\phi_2 - \Delta z \cos\phi_2) + \\ &+ a_1^2 + a_2^2 \\
        B^{\perp}(\phi_2) =& \; 2\Delta x a_1 \\
        C^{\perp}(\phi_2) =& \; 2\Delta y a_1 + 2a_1 a_2 \sin\phi_2 \\
    \end{split}
    \label{eq:inductance_ABC_O}
\end{equation}
Eqs.~(\ref{eq:inductance_rings_P}) and (\ref{eq:inductance_rings_O}) are used further to numerically test the accuracy of approximations.

\section{Recurrence relations for coefficients behind powers of $1/R$}\label{sec:recurrence}

Now let us notice that the non-quasistatic inductance defined by Eq.~(\ref{eq:inductance_loops}) as a function of the vector $\bm{R}$, which connects the circular loop centers, satisfies the Helmholtz equation
\begin{equation}
    \left(\nabla^2 + k^2\right)L_{12}(\bm{R}) = 0
    \label{eq:Helm_L}
\end{equation}
The boundary conditions are $L|_{R\rightarrow\infty}=0$ and $L|_{R=0} = L^{(0)}_{12}$, with the latter inductance being a constant for parallel loops (the self-inductance in the case $a_1=a_2$) and $0$ for orthogonal loops.

Assuming the coordinate vector to be represented in spherical coordinates $\bm{R}=(R,\vartheta,\varphi)^{\mathsf{T}}$, we consider the following ansatz for $L_{12}(\bm{R})$ expansion in a region $R>R_b$ for some positive constant $R_b>0$:
\begin{equation}
    L_{12}(\bm{R}) = e^{ikR} \sum_{n=1}^\infty \frac{\mathcal{L}_n(\vartheta,\varphi)}{(kR)^n}
    \label{eq:L_ansatz}
\end{equation}
Substitution of the latter equation into the Helmholtz equation (\ref{eq:Helm_L}) and grouping terms with the same powers of $1/R$ yields the recurrence relation on $\mathcal{L}_n$ coefficients:
\begin{equation}
    -2in\mathcal{L}_{n+1} + n(n-1)\mathcal{L}_n + \Delta_{\vartheta,\varphi}\mathcal{L}_n = 0
    \label{eq:Ln_recurrence}
\end{equation}
where $\Delta_{\vartheta,\phi}$ is the angular part of the Laplace operator. This recurrence relation allows one to successively find higher orders of expansion based on the first term.

For forward recurrence in Eq.~(\ref{eq:Ln_recurrence}) terms grow linearly relative to $n$
\begin{equation}
    |\mathcal{L}_n| \sim n
\end{equation}
resulting in
\begin{equation}
    \left| \frac{\mathcal{L}_n(\vartheta,\varphi)}{(kR)^n}\right| \sim \frac{n}{(kR)^n}, \; n\rightarrow\infty
\end{equation}
which means absolute convergence for $kR>1$. Despite the absence of stability in calculation of $\mathcal{L}_n$ this fact does not prevent the use of the recurrence in practical computations since only a few terms are sufficient to get highly accurate results as demonstrated below.

\subsection{First recurrence term}

Using the first line of Eq.~(\ref{eq:inductance_loops}), for a couple of rings parallel to $XY$ plane we can explicitly express
\begin{equation}
    \bm{R}\cdot\Delta\bm{r} = \Delta b\left[ r_2\cos(\phi_2-\alpha) - a_1\cos(\phi_1-\alpha) \right]
    \label{eq:Rdr_par}
\end{equation}
where $r_2$ measures the radial distance within the second loop $S_2$ relative to its center, and $\alpha = \arctan(\Delta y/\Delta x)$. To evaluate $\mathcal{L}_1$ we combine Eqs.~(\ref{eq:inductance_loops}), (\ref{eq:f_Taylor}) and (\ref{eq:L_ansatz}) to get
\begin{equation}
\begin{split}
    \mathcal{L}^{\|}_1 = & \frac{\mu_0}{4\pi} \sum_{n=1}^{\infty}\dfrac{\left(ik\right)^{n+1}}{n!R^{n}} \iint_{S_{2}}r_{2}dr_{2}d\varphi_{2} \cdot \\  
    & 
    \cdot 
    \left\{ 
        \: \ointop_{C_{1}} \left( \bm{R} \cdot \Delta \bm{r} \right)^{n} \hat {\bm{R}} 
    \cdot 
        \left[  
            d\bm{l}_{1} \times d\hat{z}
        \right]
    \right\} 
\end{split}
\end{equation}
with $\hat{\bm{R}}=\bm{R}/R$. Next, substituting Eq.~(\ref{eq:Rdr_par}) and performing the summation using the Taylor expansion of exponential function, we run into
\begin{equation}
    \begin{split}
        \mathcal{L}^{\|}_1 = & \frac{\mu_0}{4\pi}\frac{\Delta b}{R}ika_1 \intop_0^{a_2} r_2dr_2 \intop_0^{2\pi} d\phi_2 e^{ikr_2(\Delta b/R)\cos\phi_2} \cdot\\
        & \cdot \intop_0^{2\pi} d\phi_1 e^{-ika_1(\Delta b/R)\cos\phi_1} \cos\phi_1 = \\
        = & \pi\mu_{0}ka_1a_2J_{1}\left(ka_1\sin\vartheta\right)J_{1}\left(ka_2\sin\vartheta\right)
    \end{split}
    \label{eq:L1P}
\end{equation}
From the latter equation it is seen that all $\mathcal{L}^{\|}_n$ depend on the angle $\vartheta$ only.

Analogously, for a couple of loops lying in orthogonal planes we obtain:
\begin{equation}
    \begin{split}
        \mathcal{L}_1^{\perp} = & -\pi\mu_{0}ka_1a_2 \frac{\cos\varphi\cos\vartheta}{\sqrt{1-\cos^2\varphi\sin^2\vartheta}} \times \\
        & \times J_{1}\left(ka_1\sin\vartheta\right) J_{1}\left(ka_2\sqrt{1-\cos^2\varphi\sin^2\vartheta} \right)
    \end{split}
    \label{eq:L1O}
\end{equation}
Next terms in series (\ref{eq:L_ansatz}) follow from the recurrence relation (\ref{eq:Ln_recurrence}), and can derived either manually or by means of a symbolic differentiation computer system. The second order coefficients are given in Appendix.

\section{Numerical validation and examples}\label{sec:validation}

The main goal for the development of the approximation procedures based on Eqs.~\eqref{eq:Ln_recurrence}, \eqref{eq:L_ansatz} in place of the exact calculation is to use these results for analysis of large systems (from thousands to millions and more resonators), where direct exact calculations would be too time-consuming.
In this respect, it must be expected that the exact calculation should be retained for a relatively small number of nearest neighbours for each loop, while for numerous loop pairs separated by a large enough distance approximate procedures can be used instead.

The key question, then, is the choice of the threshold distance $R_\text{t}$ between the rings, when the exact calculation can be switched to the approximate one. This question can only be answered using an integral characteristic of the entire array, as even though the calculation mismatch for each pair of remote rings is small, the integral effect, owing to the large amount of remote neighbours, might be substantial.

Another important aspect in this matter is that a suitable threshold distance $R_\text{t}$ can be different depending on the mutual orientation of the rings, and has a certain angular dependence within the array. We show below that it is indeed the case, and provide illustrative examples of the outcomes.

\subsection{Mutual inductance for two rings}

As an initial step in the required analysis, we demonstrate the applicability of the derived approximation by comparing values of the mutual inductance $L_{int}$ obtained by direct numerical integration in Eqs.~(\ref{eq:inductance_rings_P}), (\ref{eq:inductance_rings_O}) with the corresponding approximate values $L_n$, $n=1,2,3,4$, calculated by means of Eq.~(\ref{eq:L_ansatz}) with Eqs.~(\ref{eq:L1P}) and (\ref{eq:L1O}). The values of $\mathcal{L}_n$ for $n=2,3,4$ were calculated using pyTorch automatic gradient evaluation \cite{paszke2019pytorch} based on recurrence (\ref{eq:Ln_recurrence}).

\begin{figure}[t]
    \begin{center}
        \includegraphics[width=0.8\linewidth]{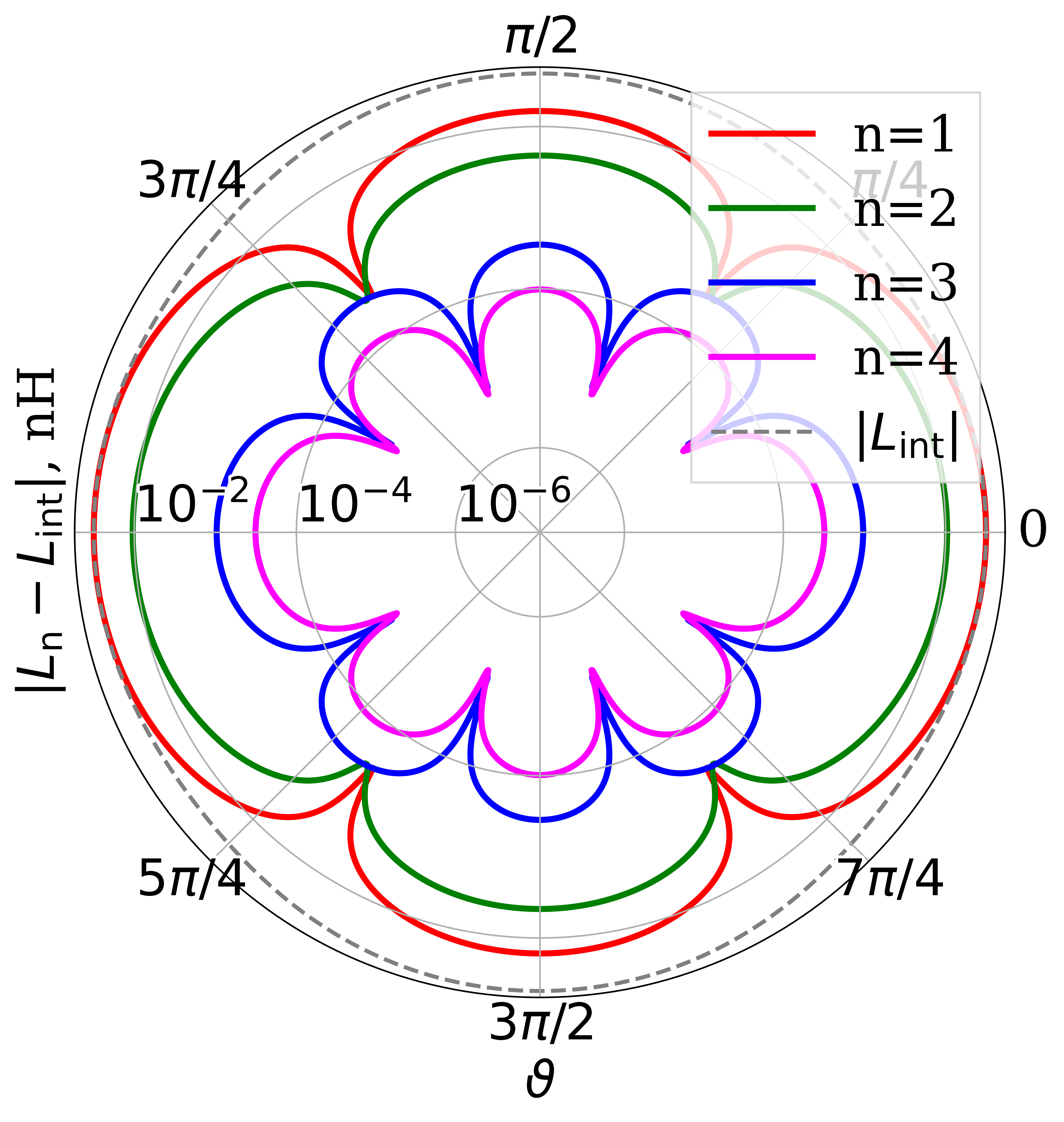}
        \caption{Angular dependence of the absolute error in inductance for four partial sums of the series (\ref{eq:L_ansatz}) for parallel loops. Parameters are $\lambda/a=20$ and $R/a=10$.}
        \label{fig:LP_ang}
    \end{center}
\end{figure}

\begin{figure}[ht]
    \begin{center}
        \includegraphics[width=0.9\linewidth]{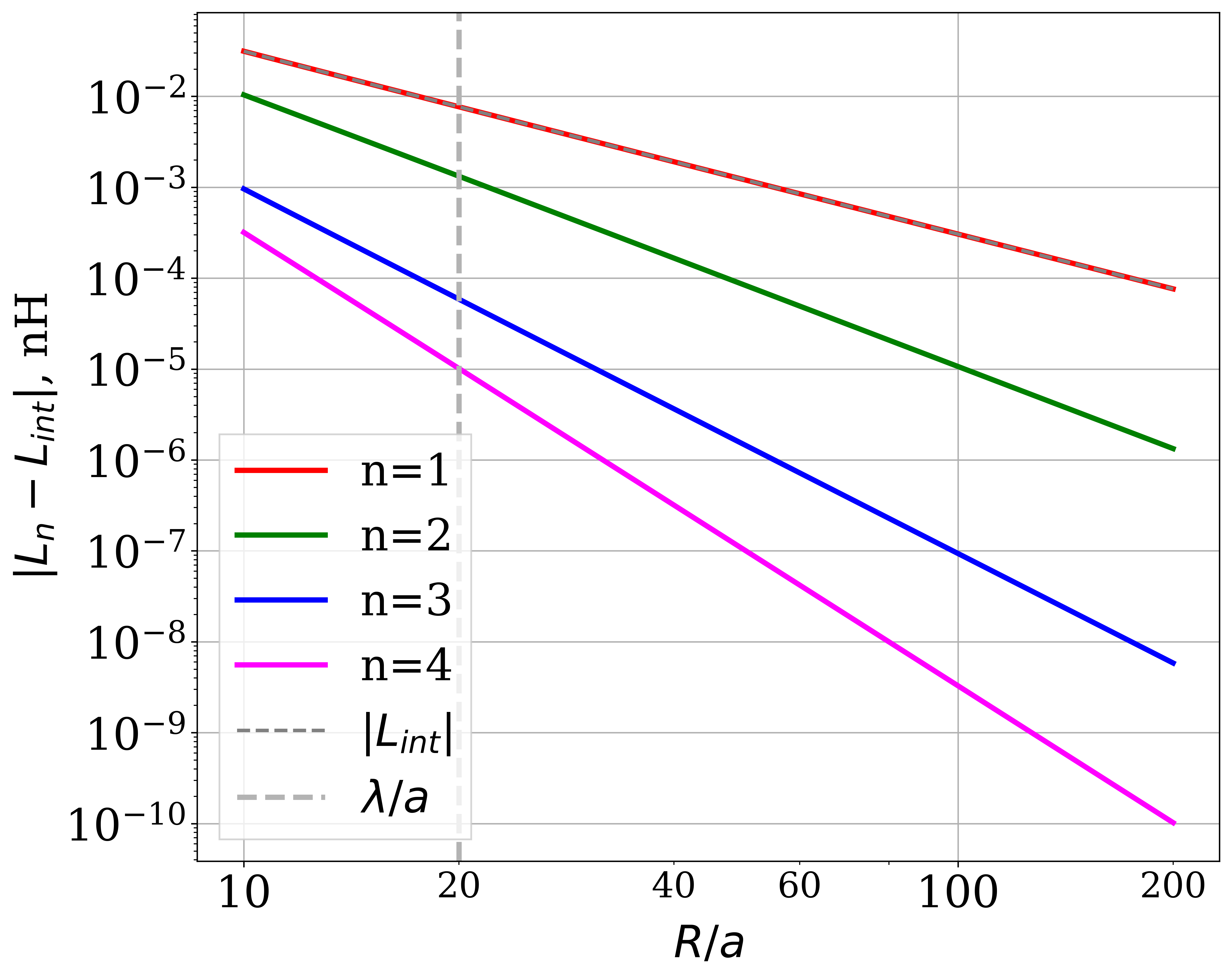}
        \caption{Dependence of the absolute error in mutual inductance from distance between centers of two parallel circular loops for four partial sums of the series (\ref{eq:L_ansatz}) in the direction $\theta=0$.}
        \label{fig:LP_dist}
    \end{center}
\end{figure}

\begin{figure}[t]
    \begin{center}
        \includegraphics[width=0.8\linewidth]{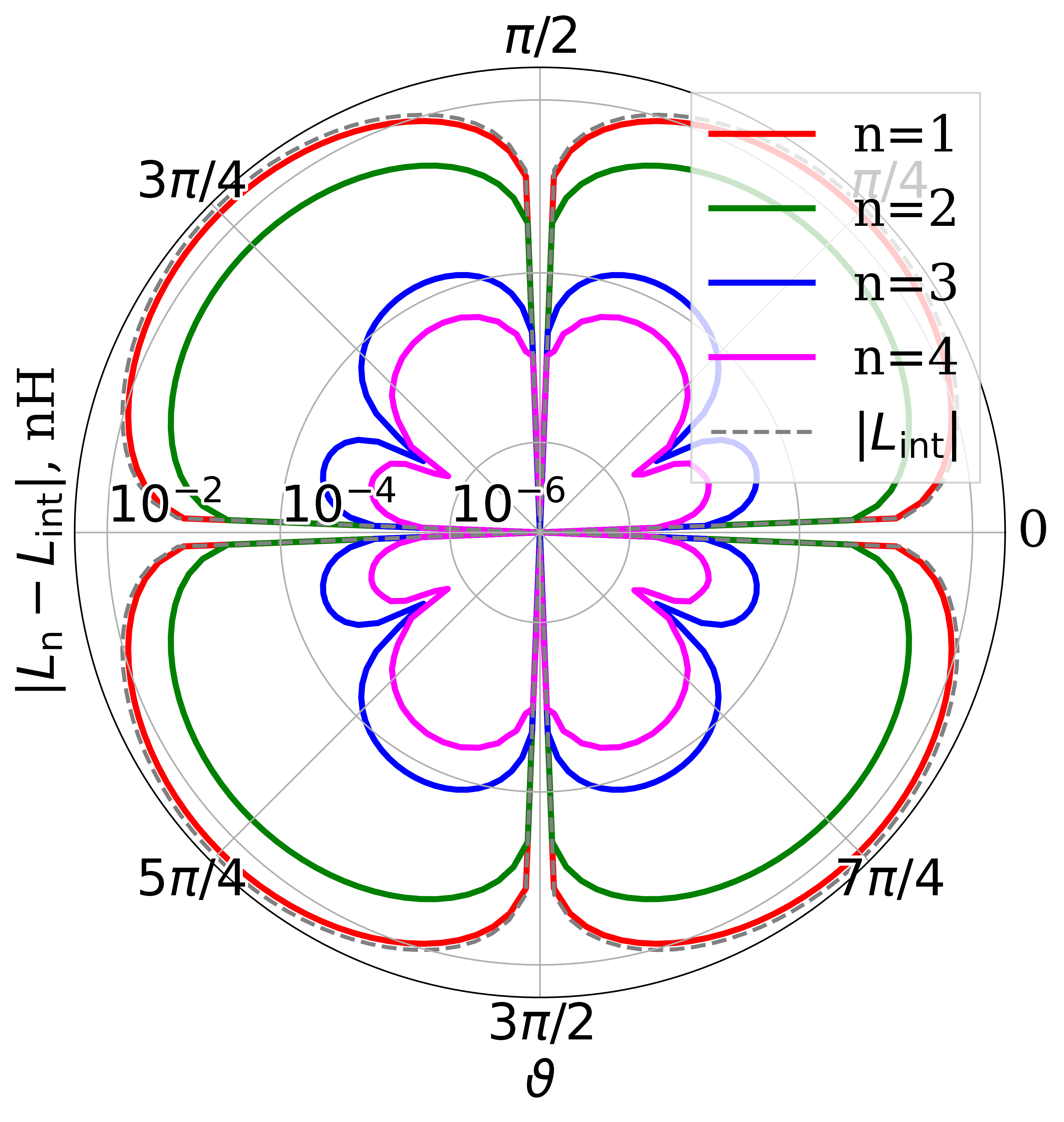}
        \caption{Angular dependence of the absolute error in inductance for four partial sums of the series (\ref{eq:L_ansatz}) for orthogonal loops. Parameters are $\lambda/a=20$ and $R/a=10$.}
        \label{fig:LO_ang}
    \end{center}
\end{figure}

\begin{figure}[ht]
    \begin{center}
        \includegraphics[width=0.9\linewidth]{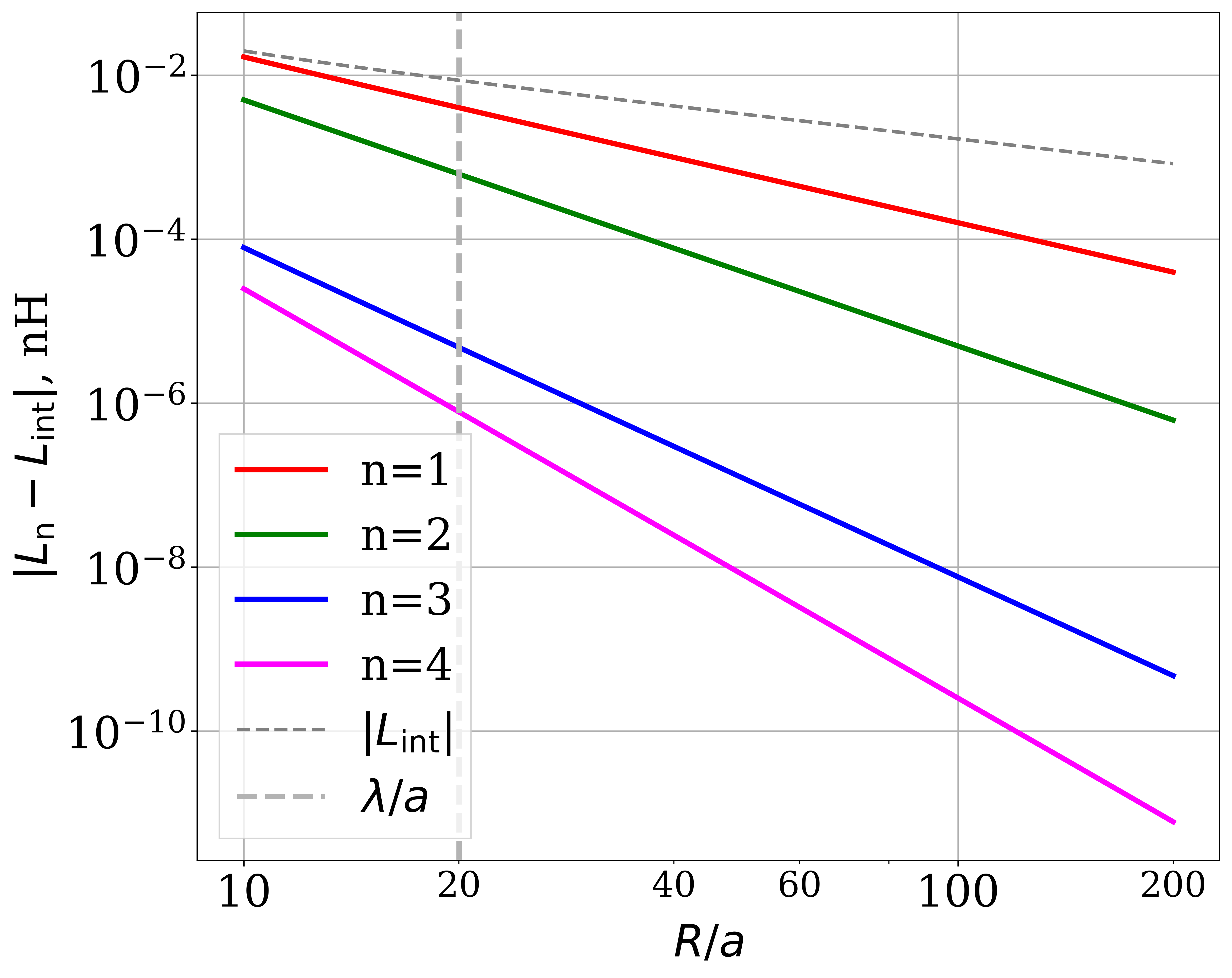}
        \caption{Dependence of the absolute error in mutual inductance from distance between centers of two parallel circular loops for four partial sums of the series (\ref{eq:L_ansatz}) in the direction $\theta=\varphi=45$.}
        \label{fig:LO_dist}
    \end{center}
\end{figure}

Fig.~\ref{fig:LP_ang} demonstrates the dependence of the absolute difference $|L_n-L_{int}|$ on the angle $\vartheta$ for two identical parallel loops with $a_1=a_2=a$ and parameters $\lambda/a=20$ and $R/a=10$. A 100\% error of the first approximation for $\vartheta=0,\pi$ is due to the fact that the Bessel functions in Eq.~(\ref{eq:L1P}) vanish for these values of the angle. Simultaneously, as one can see from the figure, the values $\vartheta=0,\pi$ identify the directions of maximum error, therefore, as a next step, we fixed $\vartheta=0$ and calculated the dependence of the absolute error from the radial distance $R$. Such dependencies are shown in Fig.~\ref{fig:LP_dist} for the chosen approximation orders. One can clearly identify the polynomial convergence in accordance with the approximation order $n$.

Figs.~\ref{fig:LO_ang}, \ref{fig:LO_dist} illustrate the same dependencies for orthogonal loops. Our experience with varying the parameters $\lambda/a$, $R/a$ and $\Delta y$ has shown that we can always obtain similar results starting from some sufficiently small value of $kR$. This fact enables efficient calculation of admittance matrices for large sets of mutually coupled loops as shown below.

\subsection{Approximations for admittance matrix}

\begin{figure}[t]
    \begin{center}
        \includegraphics[width=0.98\linewidth]{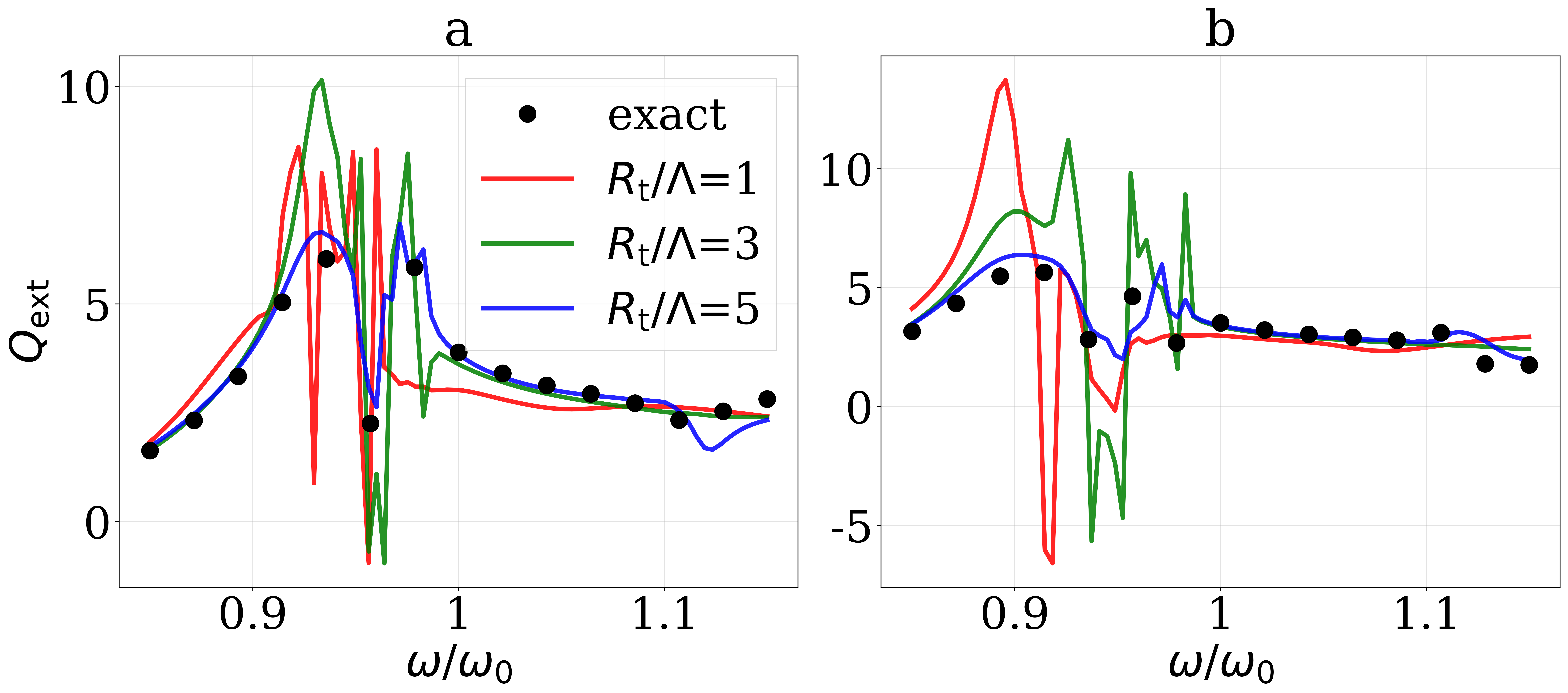}
        \caption{Extinction spectra for plane wave scattering by $10\times10\times10$ (a) and $15\times15\times15$ (b) size metamaterial cubes with fixed approximation order $n=2$ and varios values of $R_\text{t}$.}
        \label{fig:extinction}
    \end{center}
\end{figure}

Considering now applications for large systems, as an integral characteristic of the array response, we solve a problem of scattering of a plane wave by a discrete metamaterial sample and calculate the corresponding extinction efficiency. Fig.~\ref{fig:extinction} demonstrates extinction spectra around the single ring loop resonance frequency $\omega_0$ for plane wave scattering by metamaterial cubes of size $10\times10\times10$ (a) and $15\times15\times15$ (b) in the case of $n=2$ with various values of $R_\text{t}/\Lambda$, $a=0.33\Lambda$, where $\Lambda$ is a metamaterial cell size. Negative values are due to low accuracy of the approximations at small inter-loop distances. Exact values were calculated by means of the rigorous double integration expressions using Gauss-Legendre quadrature formulas. One can see that increasing $R_\text{t}$ allows reproducing exact values of extinction efficiency with acceptable accuracy and clear physical resonant behavior.

\begin{figure}[ht]
    \begin{center}
        \includegraphics[width=0.98\linewidth]{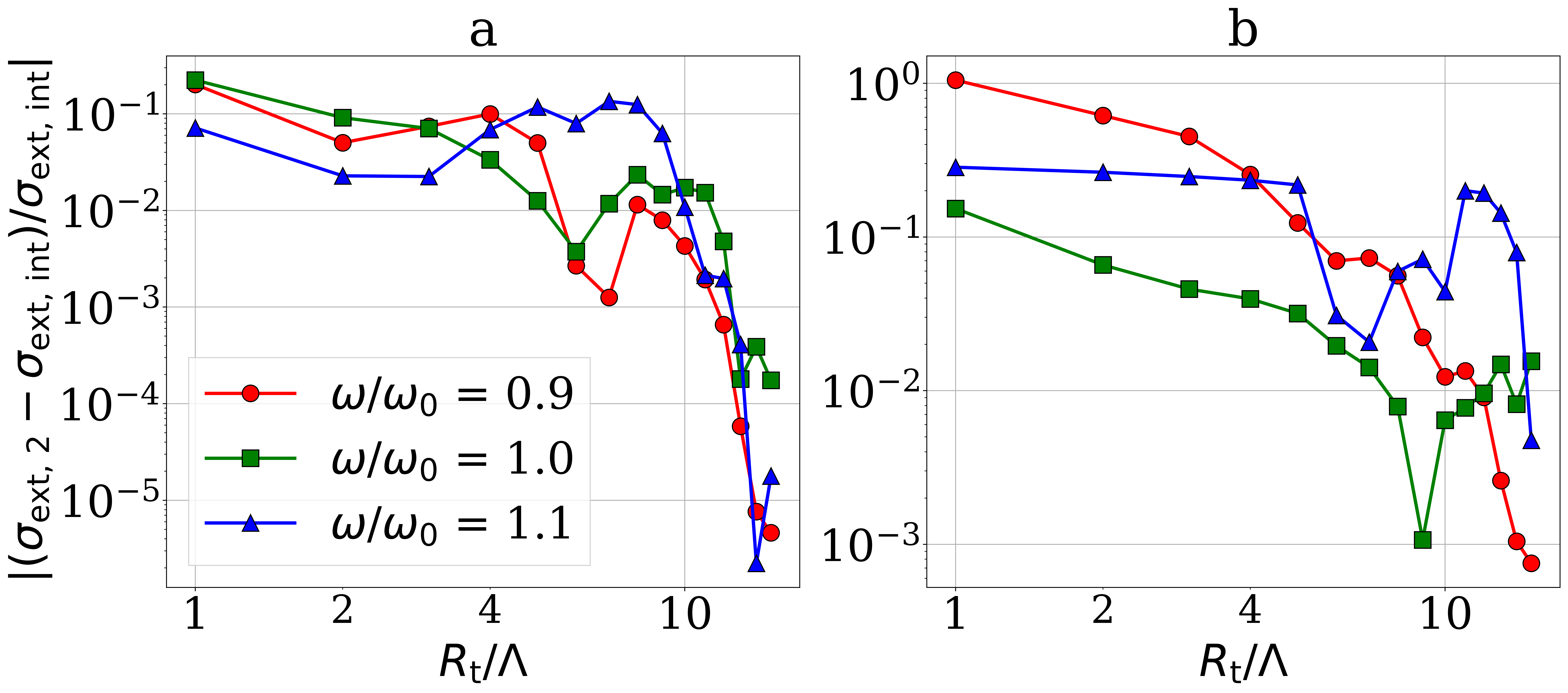}
        \caption{Dependence of the relative error for the extinction from the truncation distance $R_\text{t}$ measured in terms of metamaterial sell size $\Lambda$ when using the second order approximation in case of $10\times10\times10$ (left) and $15\times15\times15$ (right) size metamaterial cubes.}
        \label{fig:truncation_n2}
    \end{center}
\end{figure}

\begin{figure}[ht]
    \begin{center}
        \includegraphics[width=0.98\linewidth]{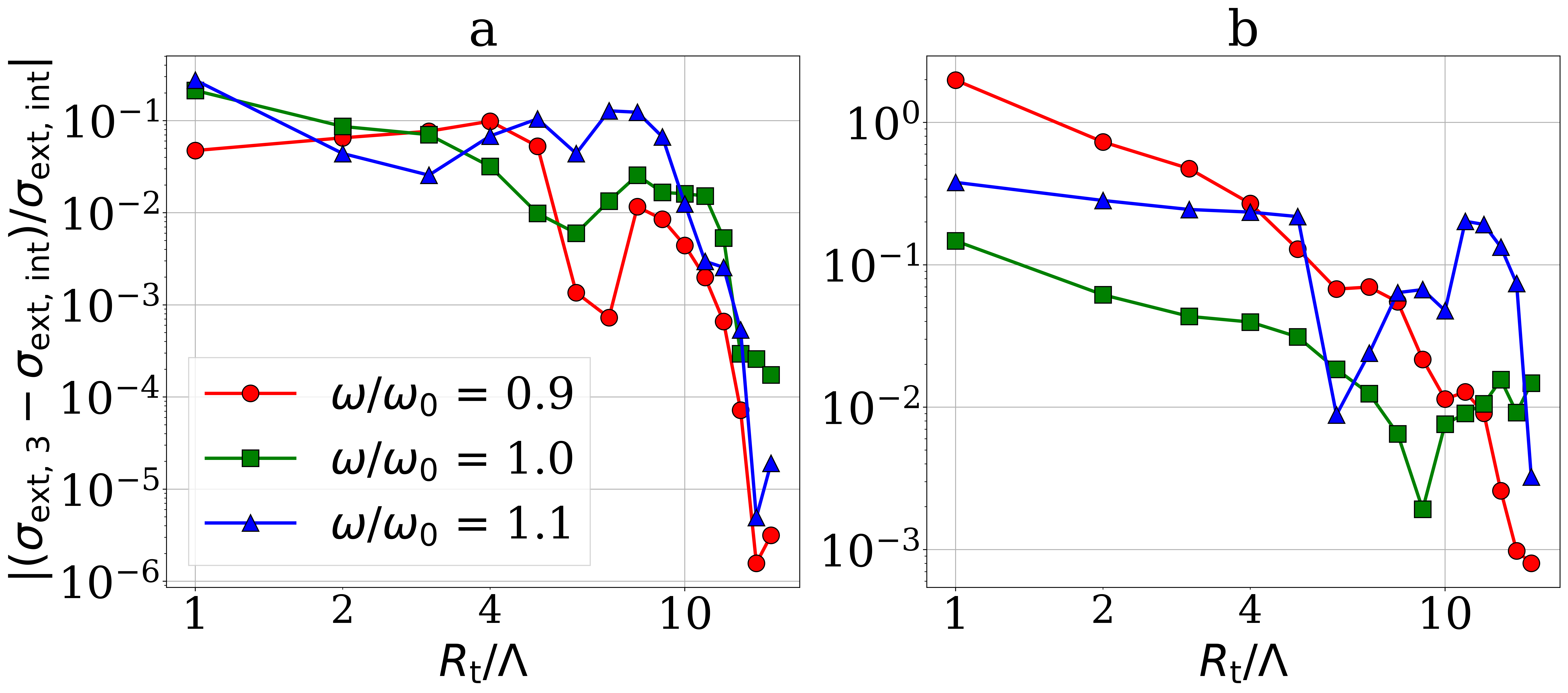}
        \caption{Dependence of the relative error for the extinction from the truncation distance $R_\text{t}$ measured in terms of metamaterial sell size $\Lambda$ when using the third order approximation in case of $10\times10\times10$ (left) and $15\times15\times15$ (right) size metamaterial cubes.}
        \label{fig:truncation_n3}
    \end{center}
\end{figure}

Figs.~\ref{fig:truncation_n2}, \ref{fig:truncation_n3} give examples of the dependence of relative extinction efficiency error on $R_\text{t}$ in the case of $n=2$ and $n=3$ order approximation for the cubes of size $10^3$ (subfigures (a)) and $15^3$ (subfigures (b)). We make here the following observations. First, an increase in the approximation order did not appreciably change the extinction accuracy and the main parameter affecting the precision is the truncation radius. Second, starting from some value of $R_\text{t}$, which depends on a sample size, we observe a sharp drop of the relative error, which allows making accurate simulations with the aid of the developed approximation.

In our Python implementation, using the approximation for inductance allowed us to accelerate extinction computations by about $4-5$ times for $R_{\text{t}}/\Lambda=3$ and by about 2-3 times for $R_{\text{t}}/\Lambda=5$.

\section{Conclusion}\label{sec:conclusion}

%
n summary, we have analyzed the problem of efficient calculation of the correct mutual inductance in the presence of retardation, whereby the distance between current loops is not sufficiently small for standard quasi-static evaluation. Such a solution is of great importance for metamaterials and metasurfaces where mutual interaction within the entire array must be taken into account, while the sample size may be of the order of a wavelength, and contain tens of thousands to millions and more meta-atoms. Moreover, we have extended our analysis to the development of suitable approximations for numerical calculation, where recurrent series are employed instead of a direct calculation, the latter being too time-consuming for large arrays. The developed calculation scheme implies controlling the method accuracy of both the mutual inductance values for each pair of loops, and the final overall physical measurable quantity such as, e.g., the scattering cross-section. These results are highly relevant for ongoing analysis of large metamaterial samples beyond the quasi-static range, providing a clear path towards practical implementation.
%

\section*{Acknowledgements}
The work was supported by the Russian Science Foundation, grant No. 22-11-00153-$\Pi$.


\bibliographystyle{IEEEtran}
\bibliography{references}

\end{document}